\documentstyle[pre,aps,preprint,epic,eepic,epsfig,rotate]{revtex} 
\begin{document}                                     
\draft
\author{Dirk Helbing}
\address{II. Institute of Theoretical Physics, University of Stuttgart,
Pfaffenwaldring 57/III, 70550 Stuttgart, Germany}
\title{Fundamentals of Traffic Flow}
\maketitle      
\begin{abstract}
From single vehicle data a number of new empirical results 
concerning the density-dependence of the velocity distribution 
and its moments as well as the characteristics of their temporal
fluctuations have been determined. These are utilized for the specification 
of some fundamental relations of traffic flow and
compared with existing traffic theories.\\[4mm]
\end{abstract}
\pagestyle{myheadings}
\markboth{D. Helbing: Fundamentals of traffic flow, PRE}
{D. Helbing: Fundamentals of traffic flow, PRE}
For the prosperity in industrialized countries, efficient traffic systems 
are indispensable. However, due to an overall increase of mobility 
and transportation during the last years, the capacity of the road
infrastructure has been reached. Some cities like Los Angeles and San Francisco
already suffer from daily traffic collapses and their environmental
consequences. About 20 percent more fuel consumption and air pollution is
caused by impeded traffic and stop-and-go traffic.
\par
For the above mentioned reasons, several models for freeway traffic
have been proposed, microscopic and macroscopic
ones (for an overview cf. Ref. \cite{Buch}). These
are used for developing traffic optimization measures
like on-ramp control, variable speed limits or re-routing systems \cite{Buch}.
For such purposes, the best models must be selected and calibrated to 
empirical traffic relations. However, some relations are difficult to
obtain, and the lack of available empirical data has caused some stagnation
in traffic modeling. 
\par
Further advances will require a close interplay between theoretical and
empirical investigations \cite{KR}. 
On the one hand, empirical findings 
are necessary to test and calibrate the various traffic models. On the 
other hand, some hardly measurable quantities and relations can be
reconstructed by means of theoretical relations. 
\par
Therefore, 
a number of fundamental traffic relations will be presented in the following.
Until now, little is known about the velocity distribution of 
vehicles, its variance or skewness. A similar thing holds for the 
functional form of the velocity-density relation 
or the variance-density relation at high densities. 
Empirical results have also been missing for the
fluctuation characteristics of the density or average velocity.
These gaps will be closed in the following. 
Although the data are varying in detail from
one freeway stretch to another, the essential conclusions are expected to be
universal.
\par
In a recent paper \cite{PRE} it has been shown that the traffic dynamics
on neighboring lanes is strongly correlated. Therefore, it is possible
to treat the total freeway cross section in an overall way. Consequently,
we will only discuss the properties of the {\em lane averages} of macroscopic
traffic quantities.
The empirical relations have been evaluated from single vehicle data
of the Dutch two-lane freeway A9 between Haarlem and Amsterdam
(for a sketch cf. Fig.~1 in Ref.~\cite{PRE}). 
These data were detected by induction loops
at discrete places $x$ of the roadway and include the 
passage times $t_\alpha(x)$,
velocities $v_\alpha(x)$, and lengths $l_\alpha(x)$ 
of the single vehicles $\alpha$. 
Consequently, it was possible to calculate the number $N(x,t)$
of vehicles which passed the cross section at place $x$
during a time interval $[t-T/2,t+T/2]$, the {\em traffic flow}
\begin{equation}
 {\cal Q}(x,t) := N(x,t) /T \, ,
\end{equation}
and the macroscopic {\em velocity moments}
\begin{equation}
 \langle v^k \rangle := \frac{1}{N(x,t)}
 \sum_{t-T/2 \le t_\alpha(x) < t + T/2} [v_\alpha(x)]^k \, .
\end{equation}
Small values of $T$ are connected with large statistical
variations of the data, but large values can cause biased
results for $k\ge 2$
\cite{PRE}. Values between 0.5 and 2 minutes seem to be the best
compromise \cite{Buch}.
The {\em vehicle densities} $\rho(x,t)$ were calculated via the 
theoretical flow formula
\begin{equation}
 {\cal Q}(x,t) = \rho(x,t) V(x,t) \, .
\end{equation}
Other evaluation methods \cite{Note} are discussed in Ref.~\cite{Buch}.
\par
We start with the discussion of the grouped empirical velocity 
distribution $P(v;x,t)$ which was obtained in the usual way:
\begin{equation}
 P(v_l;x,t) := \frac{n(x,v_l,t)}{N(x,t)} \, .
\end{equation}
Here, $n(x,v_l,t)$ denotes the number of vehicles which pass the cross section
at $x$ between times $t-T/2$ and $t+T/2$ with a velocity
$v \in [v_l-\Delta/2, v_l + \Delta/2)$. The class interval length was
chosen $\Delta = 5$\,km/h. 
\par
\begin{figure}[htbp]
\unitlength8mm
\begin{center}
\begin{picture}(16,10.6)(0,-0.8)
\put(0,9.8){\epsfig{height=16\unitlength, width=9.8\unitlength, angle=-90, 
      bbllx=50pt, bblly=50pt, bburx=554pt, bbury=770pt, 
      file=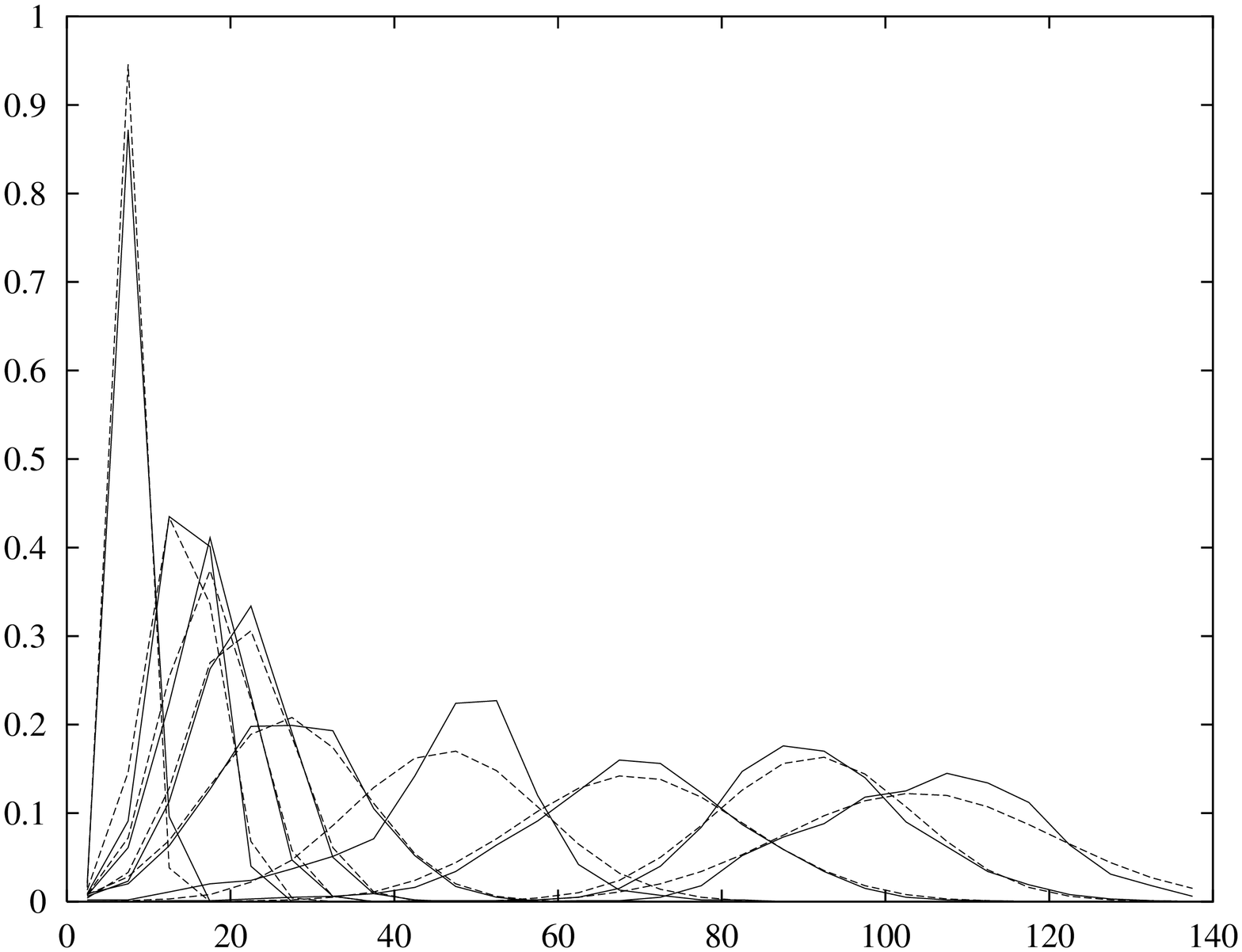}}
\put(8.7,-0.6){\makebox(0,0){\footnotesize $v$ (km/h)}}
\put(0.2,5.1){\makebox(0,0){\rotate[l]{\hbox{\footnotesize $P(v;r,t)$}}}}
\put(5.1,8){\makebox(0,0){\footnotesize $\rho=90$\,vehicles/km lane}}
\put(3.43,4.7){\makebox(0,0){\footnotesize 80}}
\put(4.24,4){\makebox(0,0){\footnotesize 70}}
\put(4.69,3.2){\makebox(0,0){\footnotesize 60}}
\put(5.15,2.5){\makebox(0,0){\footnotesize 50}}
\put(7.44,2.5){\makebox(0,0){\footnotesize 40}}
\put(8.76,2.5){\makebox(0,0){\footnotesize 30}}
\put(10.64,2.5){\makebox(0,0){\footnotesize 20}}
\put(12.35,2.2){\makebox(0,0){\footnotesize 10}}
\end{picture}
\end{center}
\caption[]{Comparison of empirical velocity distributions at different
densities (---) with frequency polygons of 
grouped Gaussian velocity distributions with
the same mean value and variance (--~--). 
A significant deviation of the empirical relations from the
respective discrete Gaussian approximations is only found at a density
of $\rho = 40$\,vehicles/km lane, where the temporal averages over
$T=2$\,min may have been too long due to rapid stop-and-go waves \cite{PRE}
(cf.\ the mysterious ``knee'' at $\rho \approx 40$\,veh/km 
in Fig.~\protect\ref{F6}).}
\label{F2}
\end{figure}

In {\em theoretical} investigations, the velocity
distribution $P(v;x,t)$ has mostly been assumed to have the Gaussian form 
\cite{Nav,PhysA,Wag}
\begin{equation}
 P_G(v;x,t) :=  \frac{1}{\sqrt{2\pi \Theta(x,t)}} \exp
 \left( - \frac{[v - V(x,t)]^2}{2\Theta(x,t)} \right) \, .
\label{gauss}
\end{equation}
Here, $V(x,t) := \langle v \rangle$ denotes the {\em average velocity}
and $\Theta(x,t) := \langle [v - V(x,t)]^2 \rangle$ the {\em velocity
variance}. Assumption (\ref{gauss}) has been made for two reasons: 
First, it allows to derive approximate fluid-dynamic traffic
equations from a gas-kinetic level of description \cite{Nav,PhysA,Wag}.
Second, analytical results for
the velocity distribution are not yet available, even for the 
stationary and spatially homogeneous case. Therefore the question is, whether
the Gaussian approximation is justified or not. Figure \ref{F2} gives
a positive answer, at least for the average velocity distribution at
small and medium densities. In particular, bimodal distributions are not
observed \cite{Note1}.
\par
\begin{figure}[htbp]
\unitlength8mm
\begin{center}
\begin{picture}(16,10.6)(0,-0.8)
\put(0,9.8){\epsfig{height=16\unitlength, width=9.8\unitlength, angle=-90, 
      bbllx=50pt, bblly=50pt, bburx=554pt, bbury=770pt, 
      file=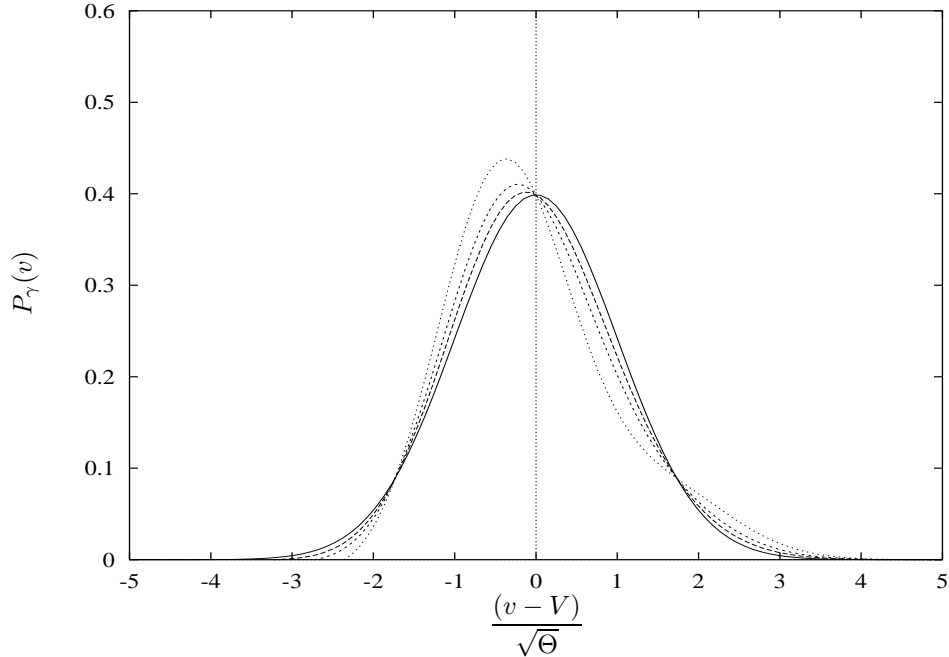}}
\put(8.7,-0.6){\makebox(0,0){\footnotesize $\displaystyle
\frac{(v-V)}{\sqrt{\Theta}}$}}
\put(0.2,5.1){\makebox(0,0){\rotate[l]{\hbox{\footnotesize $P_\gamma(v)$}}}}
\end{picture}
\end{center}
\caption[]{Velocity distributions $P_\gamma(v) 
:= \{ 1 - \gamma [3(v-V)/\Theta^{1/2} -
(v-V)^3 / \Theta^{3/2} ]/6 \} P_G(v)$ with the same average velocity
$V$ and variance $\Theta$, but different values of the skewness
$\gamma$ (---\,: $\gamma = 0$; --~--\,:
$\gamma = 1/2$; -~-~-\,: $\gamma = 1$; $\cdots$\,: 
$\gamma = 2$). Obviously, a skewness of $|\gamma| \le 0.5$ only leads to minor
changes compared to the Gaussian distribution (---).}
\label{F3}
\end{figure}
\begin{figure}[htbp]
\unitlength8mm
\begin{center}
\begin{picture}(16,10.6)(0,-0.8)
\put(0,9.8){\epsfig{height=16\unitlength, width=9.8\unitlength, angle=-90, 
      bbllx=50pt, bblly=50pt, bburx=554pt, bbury=770pt, 
      file=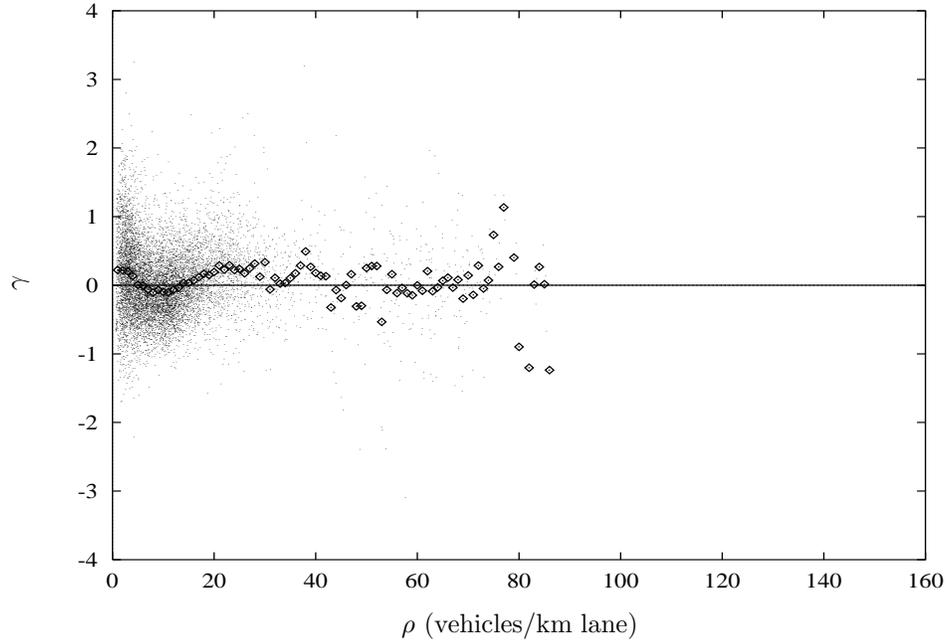}}
\put(8.7,-0.6){\makebox(0,0){\footnotesize $\rho$ (vehicles/km lane)}}
\put(0.4,5.1){\makebox(0,0){\rotate[l]{\hbox{\footnotesize $\gamma$}}}}
\end{picture}
\end{center}
\caption[]{Density-dependence of the skewness $\gamma$
($\cdot$\,: 1-minute data; $\Diamond$: respective mean values).
The large variation of the 1-minute data at low densities is due to the
small number of vehicles which pass a cross section during the time
interval $T=1$\,min, whereas the large variation of their mean values
at high densities comes from the few 1-minute data, over
which could be averaged. The 1-minute data of the skewness scatter around
the zero line (---) and mostly lie between $-0.5$ and $0.5$.} 
\label{F4}
\end{figure}

An investigation of the {\em temporal evolution} of the velocity distribution 
is difficult due to the large statistical fluctuations (which come from the
fact that only a few vehicles per velocity class
pass the observed freeway cross section during the short time period $T$).
Therefore, we will study a macroscopic (aggregated) quantity instead,
namely the temporal variation of the {\em skewness}
\begin{equation}
 \gamma(x,t) := \frac{\langle [v - V(x,t)]^3 \rangle}{[\Theta(x,t)]^{3/2}}
 = \frac{\langle v^3 \rangle - 3 \langle v \rangle \langle v^2 \rangle
 + 2 \langle v \rangle^3}{[\Theta(x,t)]^{3/2}}
 \, .
\end{equation}
This can be interpreted as a dimensionless measure of asymmetry
(cf. Fig.~\ref{F3}). Figure~\ref{F4} shows that the skewness mainly
varies between $-0.5$ and 0.5. The deviation from 0 is neither systematic
nor significant, so that the skewness is normally negligible. This indicates
that even the time-dependent velocity distribution is approximately 
Gaussian-shaped \cite{Note2}. 
\par
\begin{figure}[htbp]
\unitlength8mm
\begin{center}
\begin{picture}(16,10.6)(0,-0.8)
\put(0,9.8){\epsfig{height=16\unitlength, width=9.8\unitlength, angle=-90, 
      bbllx=50pt, bblly=50pt, bburx=554pt, bbury=770pt, 
      file=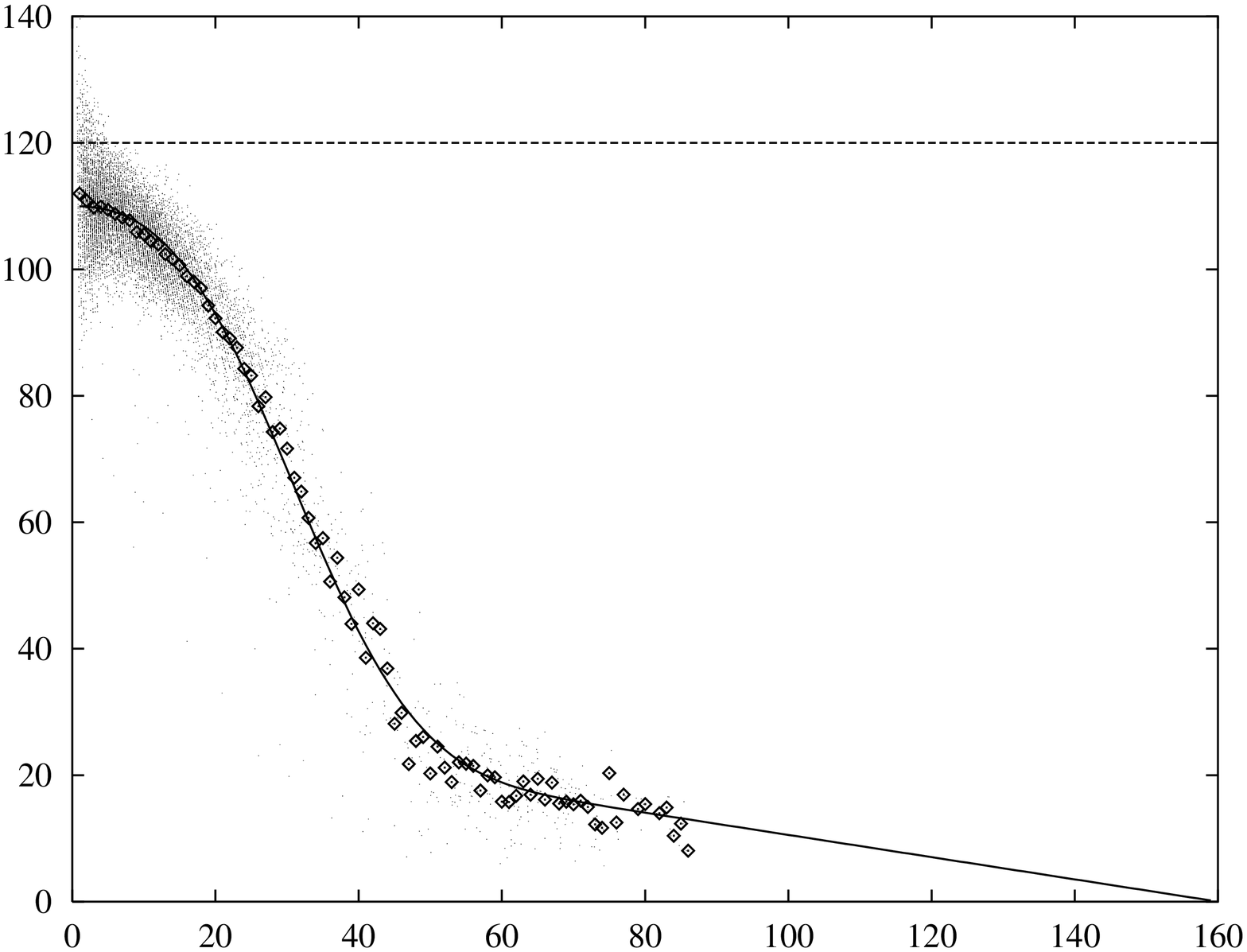}}
\put(8.7,-0.6){\makebox(0,0){\footnotesize $\rho$ (vehicles/km lane)}}
\put(0.4,5.1){\makebox(0,0){\rotate[l]{\hbox{\footnotesize $V$
(km/h)}}}}
\end{picture}
\end{center}
\caption[]{Relation between average velocity and density
($\cdot$: 1-minute data; $\Diamond$: respective mean values;
---: fit function for the equilibrium relation $V_e(\rho)$). 
The speed limit is 120\,km/h (--~--).} 
\label{F5}
\end{figure}
\begin{figure}[htbp]
\unitlength8mm
\begin{center}
\begin{picture}(16,10.6)(0,-0.8)
\put(0,9.8){\epsfig{height=16\unitlength, width=9.8\unitlength, angle=-90, 
      bbllx=50pt, bblly=50pt, bburx=554pt, bbury=770pt, 
      file=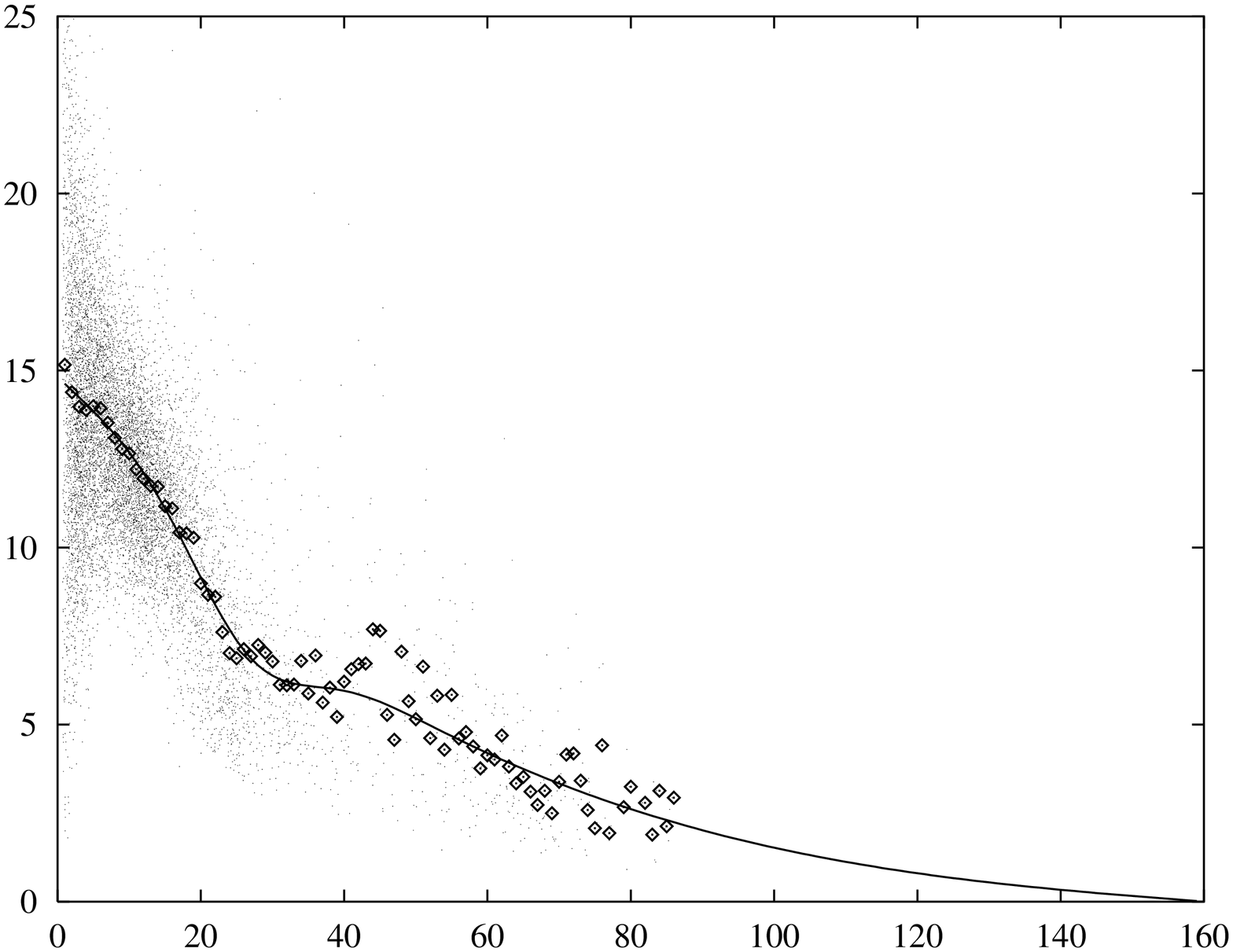}}
\put(8.7,-0.6){\makebox(0,0){\footnotesize $\rho$ (vehicles/km lane)}}
\put(0.4,5.1){\makebox(0,0){\rotate[l]{\hbox{\footnotesize $\sqrt{\Theta}$
(km/h)}}}}
\end{picture}
\end{center}
\caption[]{Density-dependence of the standard deviation $\sqrt{\Theta}$ of
the vehicle velocities 
($\cdot$: 1-minute data; $\Diamond$: respective mean values;
---: fit function for the equilibrium relation $\sqrt{\Theta_e(\rho)}$).}
\label{F6}
\end{figure}

Now it will be investigated how the average velocity $V$ and the
variance $\Theta$ depend on the vehicle density $\rho$ (cf. Figs.~\ref{F5}
and \ref{F6}). The problem is that the data for high vehicle densities are
missing. However, for computer simulations of the traffic dynamics
the corresponding functional relations need to be specified.
This can be done by means of theoretical results. 
For the average velocity and variance on freeways with speed
limits, recent gas-kinetic traffic models \cite{PhysA} 
imply the following implicit {\em equilibrium
relations} (indicated by a subscript ``$e$''), 
if the skewness is neglected (cf. Fig.~\ref{F4}):
\begin{equation}
 V_e(\rho) = V_0 - \frac{\tau(\rho) [1-p(\rho)] \rho \Theta_e(\rho)}
 {1 - \rho/\rho_{max} - \rho T_r V_e(\rho)} \, ,
\label{V}
\end{equation}
\begin{equation}
 \Theta_e(\rho) = A(\rho) [ V_e(\rho)^2 + \Theta_e(\rho) ] \, , 
 \quad \mbox{i.e.} \quad
 \Theta_e(\rho) = \frac{A(\rho)V_e(\rho)^2}{1-A(\rho)} \, .
\label{T}
\end{equation}
Herein, $V_0$ denotes the {\em average desired speed} (or free speed),
$\tau(\rho)$ is the effective density-dependent {\em relaxation time} of
acceleration maneuvers, $p(\rho)$ means the {\em probability of immediate
overtaking}. Moreover, $\rho_{max}$ denotes the {\em maximum vehicle density},
$T_r$ the {\em reaction time}, and $A(\rho)$ with 
$0 \le A(\rho) \ll 1$ the {\em relative individual velocity fluctuation}
during the time interval $\tau(\rho)$ \cite{Buch,PhysA}.
\par
According to relation (\ref{T}), the equilibrium variance vanishes when the
average velocity becomes zero. This consistency condition is not met by
all traffic models (cf. Ref. \cite{Hel}). In addition, we expect that
the average velocity vanishes at the maximum vehicle density $\rho_{max}$.
Therefore, in the limit $\rho \rightarrow \rho_{max}$ we must have
the proportionality relation
\begin{equation}
 \tau(\rho) [1-p(\rho)] \rho \frac{A(\rho) V_e(\rho)^2}{1 - A(\rho)}
  \propto 1 - \frac{\rho}{\rho_{max}} - \rho T_r V_e(\rho) \, ,
\end{equation}
the proportionality factor being $V_0$. Whereas the overtaking probability 
$p(\rho)$ is expected to vanish for $\rho \rightarrow \rho_{max}$, the
relaxation time $\tau(\rho)$ and the fluctuation parameter $A(\rho)$
are assumed to remain finite \cite{Note3}. Therefore, the {\em ansatz}
$V_e(\rho) \propto (1-\rho/\rho_{max})^\beta$ leads to $\beta = 1$ and
\begin{equation}
 V_e(\rho) = \frac{\rho_{max} - \rho}{T_r (\rho_{max})^2} \quad \mbox{for} \quad
 \rho \approx \rho_{max} \, .
\end{equation} 
This is a very interesting discovery, since many researchers believed
that the average velocity approaches the $\rho$-axis horizontally. In
addition, we find that $\Theta_e(\rho) \propto (1 -
\rho/\rho_{max})^2$ for $\rho \rightarrow \rho_{max}$.
\par
Our remaining task is to specify the parameters $\rho_{max}$ and
$T_r$. From other measurements it is known that $\rho_{max}$ lies between
160 and 180 vehicles per kilometer and lane \cite{Kuehne}. The reaction time
$T_r$ for expected events is at least 0.7 seconds \cite{May}. 
A good fit of the data results for 
\begin{equation}
 \rho_{max} = 160\,\mbox{vehicles/km lane}\,, \quad T_r = 0.8\,{\rm s} 
\end{equation}
(cf. Fig. \ref{F5}).
In addition we can conclude from (\ref{V}) that the velocity-density relation
$V_e(\rho)$ of a multi-lane freeway should start horizontally, since the
probability of overtaking $p(\rho)$ should approach the value 1 at 
very small densities $\rho \approx 0$. 
\par
\begin{figure}[htbp]
\unitlength8mm
\begin{center}
\begin{picture}(16,10.6)(0,-0.8)
\put(0,9.8){\epsfig{height=16\unitlength, width=9.8\unitlength, angle=-90, 
      bbllx=50pt, bblly=50pt, bburx=554pt, bbury=770pt, 
      file=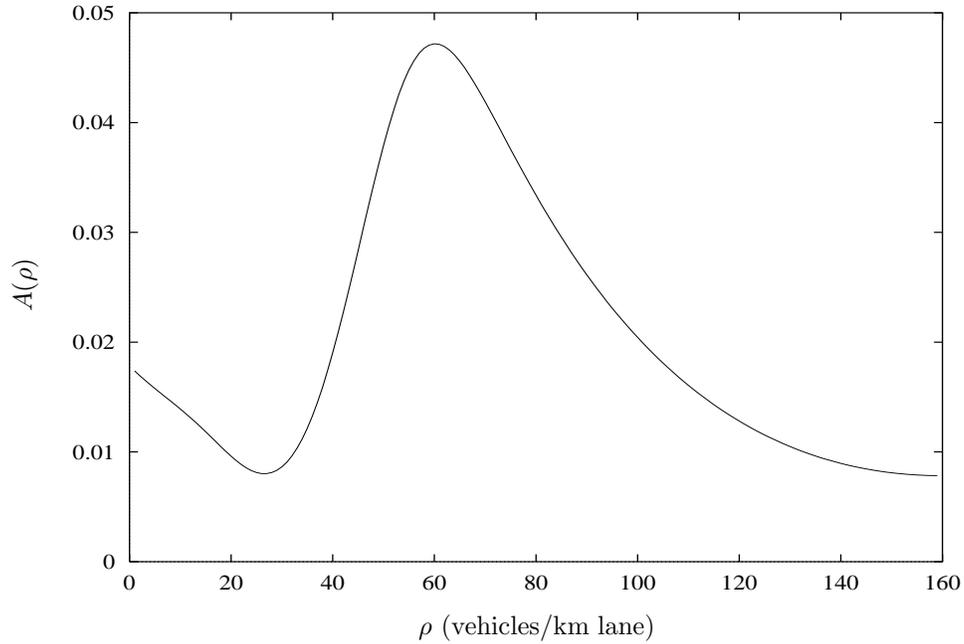}}
\put(8.7,-0.6){\makebox(0,0){\footnotesize $\rho$ (vehicles/km lane)}}
\put(0.2,5.1){\makebox(0,0){\rotate[l]{\hbox{\footnotesize $A(\rho)$}}}}
\end{picture}
\end{center}
\caption[]{Density-dependence of the fluctuation strength $A(\rho)$,
which is a measure for the relative velocity variation 
during a time interval $\tau(\rho)$. Its maximum at medium densities
indicates that velocity fluctuations are particularly large in
the region of unstable traffic flow.}
\label{F7}
\end{figure}

However, it is not only possible to reconstruct the functional forms of
the velocity-density relation $V_e(\rho)$ and the variance-density relation
$\Theta_e(\rho)$. From these we can also determine the dependence of the 
model functions $A(\rho)$ and $\tau(\rho)[1-p(\rho)]$ by means of the 
theoretical relations (\ref{V}) and (\ref{T}). The result for the
diffusion strength $A(\rho)$ is depicted in Figure~\ref{F7}.
\par
\begin{figure}[htbp]
\unitlength8mm
\begin{center}
\begin{picture}(16,10.6)(0,-0.8)
\put(0,9.8){\epsfig{height=16\unitlength, width=9.8\unitlength, angle=-90, 
      bbllx=50pt, bblly=50pt, bburx=554pt, bbury=770pt, 
      file=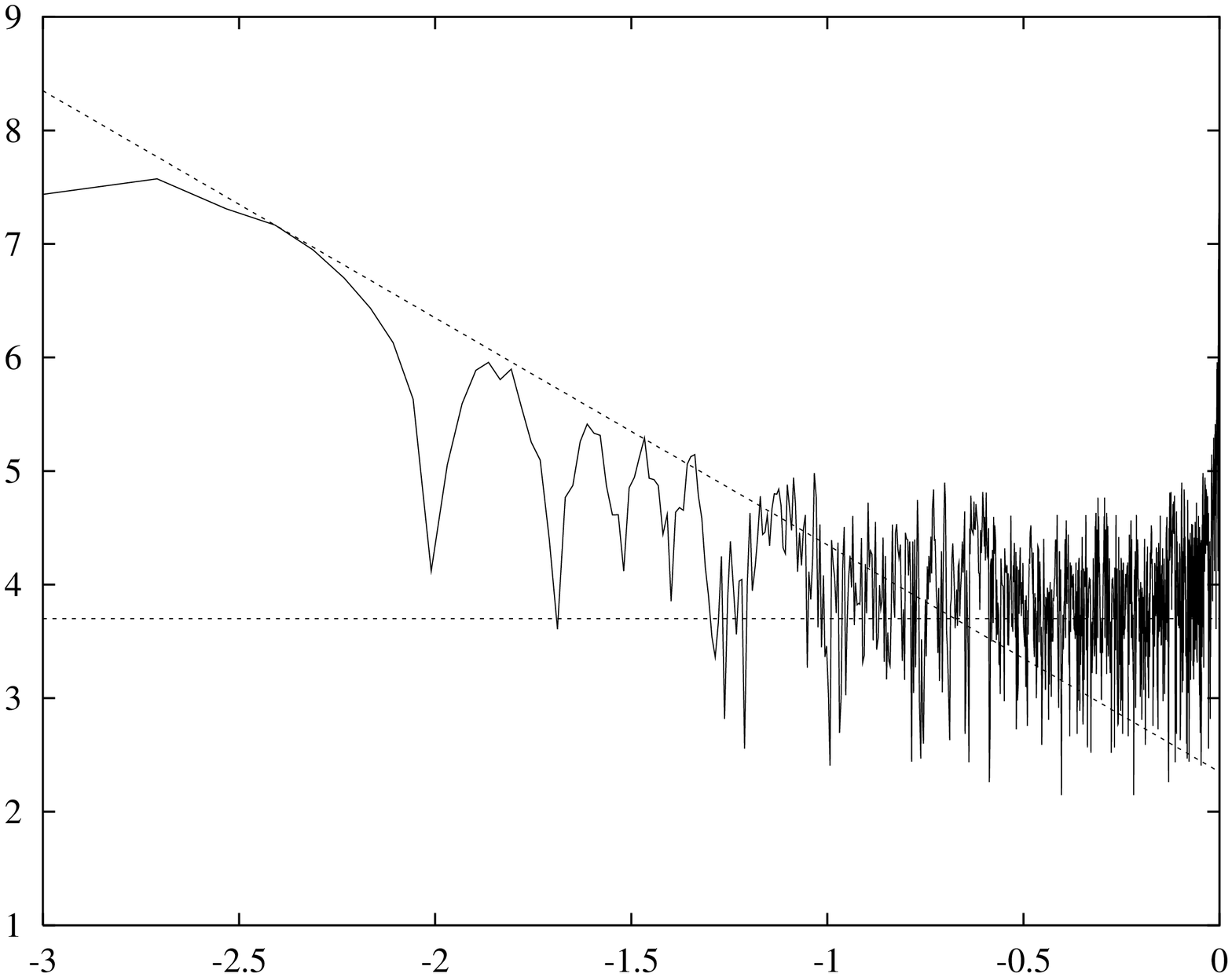}}
\put(8.7,-0.6){\makebox(0,0){\footnotesize $\displaystyle \log (\nu$ min)}}
\put(0.4,5.1){\makebox(0,0){\rotate[l]{\hbox{\footnotesize $\log 
\hat{\rho}(\nu)$}}}}
\end{picture}
\end{center}
\caption[]{The power spectrum of the time-dependent vehicle density
$\rho(x,t)$ follows a power law with exponent $\delta = 2.0$ at very
small frequencies $\nu$, but it is flat over large parts of the
frequency range, corresponding to a {\em white noise}.}
\label{F8}
\end{figure}

Finally, we will investigate the temporal fluctuations of
the empirical vehicle density $\rho(x,t)$. 
Until now, most related studies have been presented theoretical or
simulation results. It has been claimed that the power spectrum 
$\hat{\rho}(x,\nu)$ of the density $\rho(x,t)$ obeys a {\em power law}
\begin{equation}
 \hat{\rho}(x,\nu) \propto \nu^{-\delta} \, , \quad \mbox{i.e.} \quad
 \log \hat{\rho}(x,\nu) = C - \delta \log \nu \, .
\end{equation}
For $\delta$, the values 1.4 \cite{PL}, 1.0 \cite{PL1}, or 1.8
\cite{PL2} have been found. The empirical results in Figure~\ref{F8}
indicate that the exponent $\delta$ is $2.0$ at small frequencies $\nu$, 
otherwise $0.0$. Taking into account the logarithmic frequency scale,
we can conclude that the power spectrum is flat for the most
part of the frequency range. This corresponds to a {\em white noise}. 
Analogous results are found for the power spectrum of the 
average velocity $V(x,t)$ \cite{Buch}.
\par
In summary, we found that the velocity distribution is approximately
Gaussian distributed and that its skewness is negligible. 
We were able to reconstruct the velocity-density relation 
$V_e(\rho)$ and the variance-density relation $\Theta_e(\rho)$ by
means of theoretical results. This allowed the determination of some
density-dependent model parameters. The fluctuations of the vehicle
density could be approximated by a white noise, although
a power law with exponent 2.0 was found at small frequencies. 
All these results are necessary for realistic traffic simulations.

\section*{Acknowledgments}

The author is grateful to Henk Taale and the {\it Ministry of Transport,
Public Works and Water Management} for supplying the freeway data.

\clearpage
\clearpage
\clearpage
\clearpage

\begin{references}
\bibitem{Buch} D. Helbing, {\em Verkehrsdynamik. Neue physikalische
    Mo\-dellierungskonzepte} (Springer, Berlin, in preparation).
\bibitem{KR} B. S. Kerner and H. Rehborn, Phys. Rev. E {\bf 53}, 1297
and 4275 (1996).
\bibitem{PRE} D. Helbing, Empirical traffic data and their implications
for traffic modeling, submitted to Phys. Rev. E (1996).
\bibitem{Note} For example, averaging over a small stretch of length $X$ 
between $x-X/2$ and $x+X/2$ at time $t$ instead of averaging over
a time inteval $T$ at place $x$ will not exactly lead to the same
results \cite{Buch}. However, the difference in the velocity moments
is of order $\Theta/V^2$ and therefore negligible.
\bibitem{Nav} D. Helbing, Phys. Rev. E {\bf 53}, 2366 (1996).
\bibitem{PhysA} D. Helbing, Derivation and empirical validation of a
refined traffic flow model, Physica A, in print (1996).
\bibitem{Wag} C. Wagner {\it et al.}, Second order continuum traffic
flow model, Phys. Rev. E, submitted (1996).
\bibitem{Note1} Publications by W. F. Phillips [Transportation Planning and 
Technology {\bf 5}, 131 (1979)] and by R. D. K\"uhne 
[in {\em Proceedings of the 9th International Symposium on Transportation 
and Traffic Theory}, edited by I. Volmuller and R. Hamerslag (VNU Science, 
Utrecht, 1984)] have reported about bimodal distributions at
large vehicle densities. However, the reason seems to be that they chose
a large time interval $T$. Consequently, fast changes of the traffic conditions
like stop-and-go waves could produce a velocity distribution that 
appears bimodal. 
\bibitem{Note2} An exact proof of this conclusion would require a
comparison of all higher empirical velocity moments with the corresponding
relations for a Gaussian distribution. However, for many theoretical
considerations it is sufficient to know that the skewness vanishes and that
the velocity distribution is unimodal.
\bibitem{Hel} D. Helbing, Phys. Rev. E {\bf 51}, 3164 (1995).
\bibitem{Note3} This is supported by empirical observations. If $\tau(\rho)$
would diverge for $\rho \rightarrow \rho_{max}$, a traffic
jam would not be able to dissolve, once is came to rest. 
Moreover, vehicles always keep some 
minimal distance from each other, so that $\rho_{max}$ is smaller than 
the reciprocal $1/l$ of the average vehicle length $l$. This guarantees the
possibility to move.
\bibitem{Kuehne} R. K\"uhne, in {\em Highway Capacity and Level of Service}, 
edited by U. Brannolte (Balkema, Rotterdam, 1991) and unpublished material.
\bibitem{May} A. D. May, {\em Traffic Flow Fundamentals}
(Prentice Hall, Englewood Cliffs, NJ, 1990).
\bibitem{PL} T. Musha and H. Higuchi, Japanese 
Journal of Applied Physics {\bf 17}, 811 (1978).
\bibitem{PL1} K. Nagel and H. J. Herrmann, Physica A {\bf 199}, 254 (1993);
K. Nagel and M. Paczuski, Phys. Rev. E {\bf 51}, 2909 (1995);
X. Zhang and G. Hu, Phys. Rev. E {\bf 52}, 4664 (1995);
M. Y. Choi and H. Y. Lee, Phys. Rev. E {\bf 52}, 5979 (1995).
\bibitem{PL2} S. Yukawa and M. Kikuchi, Journal of the Physical Society of
  Japan {\bf 65}, 916 (1996).
\end{references}
\end{document}